\documentclass[prd,preprint,superscriptaddress,showpacs,byrevtex]{revtex4}
\usepackage{bm}
\def\fsl#1{\setbox0=\hbox{$#1$}           
   \dimen0=\wd0                                 
   \setbox1=\hbox{/} \dimen1=\wd1               
   \ifdim\dimen0>\dimen1                        
      \rlap{\hbox to \dimen0{\hfil/\hfil}}      
      #1                                        
   \else                                        
      \rlap{\hbox to \dimen1{\hfil$#1$\hfil}}   
      /                                         
   \fi}                                         %
\newcommand{\be}{\begin{equation}}
\newcommand{\ee}{\end{equation}}
\newcommand{\bea}{\begin{eqnarray}}
\newcommand{\eea}{\end{eqnarray}}
\newcommand{\beq}{\begin{equation}}
\newcommand{\eeq}{\end{equation}}
\newcommand{\beqs}{\begin{eqnarray}}
\newcommand{\eeqs}{\end{eqnarray}}

\begin{document}
\title{ Momentum Sum Rule Violation in QCD at High Energy Colliders and Confinement }
\author{Gouranga C Nayak }\thanks{G. C. Nayak was affiliated with C. N. Yang Institute for Theoretical Physics in 2004-2007.}
\affiliation{ C. N. Yang Institute for Theoretical Physics, Stony Brook University, Stony Brook NY, 11794-3840 USA}
\date{\today}
\begin{abstract}
Momentum sum rule in QCD is widely used at high energy colliders. Although the exact form of the confinement potential energy is not known but the confinement potential energy at large distance $r$ can not rise slower than ${\rm ln}(r)$. In this paper we find that if the confinement potential energy at large distance $r$ rises linearly with $r$ (or faster) then the momentum sum rule in QCD is violated at the high energy colliders if we consider the hadron size as infinite. For finite size hadron we find that the momentum sum rule in QCD is violated at the high energy colliders for any form of the confinement potential energy.
\end{abstract}
\maketitle
\pagestyle{plain}

\pagenumbering{arabic}

\section{Introduction}

After the discovery of the asymptotic freedom  \cite{gwp} in the renormlized QCD  \cite{tv} there has been lot of progress in the perturbative QCD (pQCD) calculation of the partonic scattering cross section at the high energy colliders. In spite of this success, the partonic scattering cross section we calculate in the renormalized pQCD is not a physical observable because we do not directly experimentally measure partons but we experimentally measure hadrons. This is because partons are confined inside hadrons. Hence how the partons form hadrons needs to be studied.

The formation of hadrons from partons is a long distance phenomena where the renormalized pQCD is not applicable but the renormalzied non-perturbative QCD is applicable. However, the analytical method to study the renormalized non-perturbative QCD is not known. Hence we depend on the numerical lattice QCD method to study the confinement of partons inside the hadron within the renormalized QCD.

At high energy colliders we can apply the renormalized pQCD method to calculate the partonic level cross section ${\hat \sigma}$ for the high momentum transfer partonic level scattering processes because the coupling becomes small at high momentum transfer. Using the factorization theorem \cite{fal,nal,nnkl} the partonic level cross section ${\hat \sigma}$ is folded with the non-perturbative parton distribution functions inside the hadron and non-perturbative parton to hadron fragmentation functions to calculate the hadron production cross section $\sigma$ at the high energy colliders which is experimentally measured.

Since the parton distribution function inside the hadron and the parton to hadron fragmentation function are non-perturbative quantities in the renormalized QCD they have not been studied analytically and hence we (indirectly) extract them from the experiments. The quark, antiquark and gluon distribution functions $q(\zeta,Q^2)$, ${\bar q}(\zeta,Q^2)$ and $g(\zeta,Q^2)$ inside the hadron $H$ are measured at some momentum transfer scale $Q_0$ and then evolved to another momentum transfer scale $Q$ by using DGLAP equation \cite{dl} where $\zeta$ is the momentum fraction of the parton with respect to the hadron. Although the range of $\zeta$ goes from 0 to 1 but the experiments do not measure the parton distribution functions at all values of $\zeta$.

For this reason various sum rules of partons and the DGLAP equation play important roles to determine the parton distribution functions at many values of $\zeta$ and at many values of $Q^2$ even if the experimental data are available at limited values of $\zeta$ and at limited values of $Q^2$. For example, since the gluon structure function measurements are limited than the quark structure function measurements, one can use the momentum sum rule in QCD at the high energy colliders to constrain the gluon structure function from the measured quark structure functions. Hence various sum rules have been very useful to predict the parton distribution functions (PDF's) at many values of $\zeta$ and $Q^2$ in various PDF sets such as CTEQ \cite{ct}, GRV \cite{gr} and MRST \cite{mr} etc. even if the experimental data are available at limited values of $\zeta$ and at limited values of $Q^2$

For the hadron with momentum $p_H$ at high energy colliders the momentum sum rule in QCD predicts that the sum of the momentum of all the partons inside the hadron is equal to the momentum $p_H$ of the hadron. The momentum sum rule in QCD at high energy colliders is mathematically expressed as \cite{cl,coln}
\bea
1=\sum_q \int_0^1 d\zeta~\zeta~q(\zeta,Q^2) + \sum_{\bar q} \int_0^1 d\zeta~\zeta~ {\bar q}(\zeta,Q^2)+\sum_g \int_0^1 dx ~\zeta~g(\zeta,Q^2)
\label{msr}
\eea
where $\sum_q,~\sum_{\bar q},~\sum_g$ means sum over all the quarks, antiquarks and gluons inside the hadron $H$. The momentum $<{\hat p}_q>$ of a quark, $<{\hat p}_{\bar q}>$ of an antiquark and $<{\hat p}_g>$ of a gluon inside the hadron are given by
\bea
<{\hat p}_q>=\int_0^1 d\zeta~\zeta~ { q}(\zeta,Q^2),~~~~~<{\hat p}_{\bar q}>=\int_0^1 d\zeta~\zeta~ {\bar q}(\zeta,Q^2),~~~~~~~<{\hat p}_g>=\int_0^1 d\zeta~\zeta~ { g}(\zeta,Q^2)\nonumber \\
\label{mo}
\eea
where ${\hat p}$ is the momentum operator and the expectation value inside the hadron $H$ with the color singlet (physical) state $|H>$ is defined by
\bea
<{\hat p}>=<H|{\hat p}|H>.
\label{st}
\eea
Since $p_H$ is the momentum of the hadron at high energy colliders, the momentum sum rule in eq. (\ref{msr}) in QCD at high energy colliders can be written as
\bea
p_H= \sum_{ q} <{\hat p}_{ q}>+\sum_{\bar q} <{\hat p}_{\bar q}>+\sum_{g} <{\hat p}_{g}>
\label{mtr}
\eea
where the momentum $<{\hat p}_q>$ of a quark, $<{\hat p}_{\bar q}>$ of an antiquark and $<{\hat p}_g>$ of a gluon inside the hadron are given by eq. (\ref{mo}).

Hence the momentum sum rule in QCD at high energy colliders as given by eq. (\ref{mtr}) states that the sum of the momentum of all the partons inside the hadron is equal to the momentum of the hadron at the high energy colliders. This, however, is not true which we will show in this paper because the quarks, antiquarks and gluons in QCD are confined inside the hadron and the hadron has finite size.

We find in this paper that if the confinement potential energy at large distance $r$ rises linearly with $r$ (or faster) then the momentum sum rule in QCD at high energy colliders as given by eq. (\ref{mtr}) is violated if the hadron size is infinite. Similarly we find that if the hadron size is finite then the momentum sum rule in QCD at high energy colliders as given by eq. (\ref{mtr}) is violated for any form of the confinement potential energy.

If one considers the hadron size as infinite and if the confinement potential energy at large distance $r$ rises linearly with $r$ (or faster) then we find in this paper that
\bea
p_H= \sum_{ q} <{\hat p}_{ q}>+\sum_{\bar q} <{\hat p}_{\bar q}>+\sum_{g} <{\hat p}_{g}>+ <{\hat p}_{\rm flux}>
\label{fni}
\eea
which does not agree with eq. (\ref{mtr}) where the momentum flux $<{\hat p}_{\rm flux}>$ of the partons is non-zero due to confinement of partons inside the hadron. The gauge invariant definition of the momentum flux $p_{\rm flux}$ in the Yang-Mills theory is given by eq. (\ref{ymx}).

Similarly if the size of the hadron is finite then eq. (\ref{fni}) is valid for any form of the confinement potential energy where the momentum flux $<{\hat p}_{\rm flux}>$ of the partons is non-zero. Hence we find that the non-zero momentum flux $<{\hat p}_{\rm flux}>$ in QCD should be added to the momenta of the quarks, antiquarks and gluons inside the hadron to reproduce the momentum of the hadron.

We will provide a derivation of eq. (\ref{fni}) in this paper.

The paper is organized as follows. In section II we briefly review the conservation of momentum and the vanishing momentum flux in the Dirac-Maxwell theory. In section III we discuss the non-perturbative QCD at long distance and the classical Yang-Mills theory. In section IV we find that if the confinement potential energy at large distance $r$ rises linearly with $r$ (or faster) then the momentum sum rule in QCD at high energy colliders as given by eq. (\ref{mtr}) is violated if the hadron size is infinite. In section V we discuss the momentum sum rule violation for parton distributions of color singlet hadron. In section VI we find that if the hadron size is finite then the momentum sum rule in QCD at high energy colliders as given by eq. (\ref{mtr}) is violated for any form of the confinement potential energy. Section VII contains conclusions.

\section{ Conservation of Momentum and Vanishing Momentum Flux in Dirac-Maxwell Theory }

Note that unlike QCD there is no confinement in QED. Hence in order to see the effect of confinement on momentum conservation in QCD it is useful to compare it with the momentum conservation in QED where there is no confinement. For this reason we will briefly review the vanishing momentum flux and the conservation of momentum in the Dirac-Maxwell theory in this section.

In gauge theory the conservation of gauge invariant momentum from the first principle can be described by using the gauge invariant Noether's theorem obtained by using Lorentz transformation plus local gauge transformation \cite{le,lc}. From the continuity equation of the gauge invariant energy-momentum tensor obtained from the gauge invariant Noether's theorem in the Dirac-Maxwell theory we find \cite{le}
\bea
&&\frac{d [{p}^i_{EM}+{ p}^i_e]}{dt}=- \frac{1}{2}\int d^3x ~\partial_k [ ~2E^i(x) E^{k}(x)-\delta^{ik}[{\vec E}^2(x) -{\vec B}^2(x)]\nonumber \\
&&+ {\bar \psi}(x)[\gamma^k  (i{\overrightarrow \partial}^i -eA^i(x)) -\gamma^k (i{\overleftarrow \partial}^i +eA^i(x)) ] \psi(x)~]
\label{cp}
\eea
where $\psi(x)$ is the Dirac field of the electron, ${\vec E}~({\vec B})$ is the electric (magnetic) field and the gauge invariant momentum ${\vec p}_{\rm EM}$ of the electromagnetic field $A^\nu(x)$ and the gauge invariant momentum of the electron are given by
\bea
{\vec p}_{\rm EM}=\int d^3x ~{\vec E}(x) \times {\vec B}(x),~~~~~~~~~~~~{\vec p}_e = \int d^3x~ \psi^\dagger(x) [i{\vec \partial} -e{\vec A}(x)] \psi(x).
\label{eem}
\eea
For the boundary surface at infinity the $A^\lambda(t,r)$ decays like $\frac{1}{r}$ which means vanishing momentum flux in the Dirac-Maxwell theory
\bea
\int d^3x \partial_k [ 2E^i(x) E^{k}(x)-\delta^{ik}[{\vec E}^2(x) -{\vec B}^2(x)]+ {\bar \psi}(x)[\gamma^k  (i{\overrightarrow \partial}^i -eA^i(x)) -\gamma^k (i{\overleftarrow \partial}^i +eA^i(x)) ] \psi(x)]=0.\nonumber \\
\label{vmf}
\eea
Using eq. (\ref{vmf}) in (\ref{cp}) we find
\bea
&&\frac{d [{\vec p}_{EM}+ {\vec p}_e]}{dt}=0
\label{mp}
\eea
which proves that the sum of the momentum of the field plus the particle is conserved in the Dirac-Maxwell theory if the entire infinite volume is considered.

\section{ Non-Perturbative QCD at Long Distance and The Classical Yang-Mills Theory }\label{pbd}

Note that the QED is the quantum theory of the classical electrodynamics. Similarly the QCD is the quantum theory of the classical Yang-Mills theory. If the boundary surface is at the finite distance then the boundary surface term is not zero both in QED and in QCD.

In the previous section we saw that if the boundary surface is at infinite distance then the boundary surface term is zero in the Dirac-Maxwell theory. In order to study momentum conservation equation in QCD we need to know if the boundary surface term at infinite distance is zero or non-zero in QCD if one takes the hadron size as infinite. Note that in reality the size of the hadron is finite where the boundary surface term is non-zero which violate the momentum sum rule in QCD at high energy colliders for any form of the confinement potential energy (see section VI for more details).

The infinite distance is a long distance in QCD where the renormalized pQCD is not applicable but the renormalized non-perturbative QCD is applicable. In order to know how the fields behave at long distance in renormalized QCD we need to solve the renormalized non-perturbative QCD. However, we do not know the analytical method to study the renormalized non-perturbative QCD at long distance and the lattice QCD has not studied these boundary surface terms at the large distances.

The QED becomes non-perturbative at short distance and the renormalized QCD becomes perturbative at short distance. Similarly the QED becomes perturbative at long distance and the renormalized QCD becomes non-perturbative at long distance. Note that the exact form of the Coulomb potential produced by the electron can be derived by solving the classical Maxwell theory. Similarly the exact form of the Coulomb potential can be derived from the QED at the long distance. At short distance the form of the potential in QED becomes different from the Coulomb form due to quantum loops (vacuum polarization). However, at long distance the QED predicts the Coulomb form of the potential. One can include the quantum loops (vacuum polarization) in QED at long distance but such corrections will be zero at distance infinity. Hence as far as the boundary surface at infinity is concerned the QED at long distance predicts the Coulomb potential.

This analogy of the QED at the long distance and the classical Maxwell theory can be carried to renormalized non-perturbative QCD at long distance and the classical Yang-Mills theory because the Yang-Mills theory was discovered by making analogy with the Maxwell theory by extending the U(1) gauge group to SU(3) gauge group \cite{pyang}. For example in analogy to the covariant derivative $D_\nu =\partial_\nu +ie A_\nu$ in the U(1) gauge theory the covariant derivative in the Yang-Mills theory is taken to be $D_\nu =\partial_\nu -igT^c A_\nu^c$ by extending U(1) gauge group to SU(3) gauge group \cite{pyang}. Similarly in analogy to the Maxwell field tensor $[D_\mu,D_\lambda] =-ie F_{\mu \lambda}$ in the U(1) gauge theory the formula of the Yang-Mills field tensor is obtained from $[D_\mu,D_\lambda] =igT^cF_{\mu \lambda}^c$ by extending U(1) gauge group to SU(3) gauge group \cite{pyang}. Similarly in analogy to the Maxwell field lagrangian density ${\cal L} =-\frac{ F_{\mu \lambda}F^{\mu \lambda}}{4}$ in the U(1) gauge theory the Yang-Mills field lagrangian density is taken to be ${\cal L} =-\frac{ F_{\mu \lambda}^cF^{\mu \lambda c}}{4}$ by extending U(1) gauge group to SU(3) gauge group \cite{pyang}. Similarly in analogy to the Maxwell potential $A^\nu$ obtained from the pure gauge potential $A^\nu_{pure}$ in U(1) gauge theory, the Yang-Mills potential $A_\nu^c$ is obtained from the SU(3) pure gauge potential $A^{\nu c}_{pure}$ in Yang-Mills theory \cite{ppn,pcn}. Hence we find that the Yang-Mills theory is obtained by making analogy with the Maxwell theory by extending U(1) gauge group to SU(3) gauge group \cite{pyang,ppn,pcn}.

We have seen earlier in this section that as far as the boundary surface at infinite distance is concerned the QED at long distance predicts the Coulomb potential which is the same potential predicted by the classical Maxwell theory. Since the Yang-Mills theory is obtained by making analogy with the Maxwell theory by extending U(1) gauge group to SU(3) gauge group \cite{pyang,ppn,pcn} we find that as far as the boundary surface at infinite distance is concerned the renormalized non-perturbative QCD at infinite distance predicts the same potential that is predicted by the classical Yang-Mills theory. Hence we find that as far as the boundary surface at infinity is concerned a non-vanishing boundary surface term in the classical Yang-Mills theory means a corresponding non-vanishing boundary surface term in the renormalized non-perturbative QCD at infinite distance.

\section{ Momentum Sum Rule Violation in QCD at High Energy Colliders }

The electric charge $e$ of the electron is constant in the classical Maxwell theory but the fundamental color charge $q^c(t)$ of the quark is time dependent in the classical Yang-Mills theory where $c=1,2,...,8$ is the color index \cite{pcn}. The form of the Yang-Mills potential (the color potential) $A_\nu^c(x)$ produced by the color charge $q^c(t)$ of the quark in the classical Yang-Mills theory is given in \cite{ppn} which is different from the form $\frac{1}{r}$.

For a system containing quarks plus antiquarks plus the Yang-Mills potential $A_\nu^c(x)$ we find from the gauge invariant Noether's theorem in the Yang-Mills theory \cite{lc}
\bea
&&\frac{d {p}^i_{YM}}{dt}+\sum_{ q}\frac{d { p}^i_q}{dt}+\sum_{\bar q}\frac{d { p}^i_{\bar q}}{dt}=-  \frac{1}{2}\int d^3x ~\partial_k [ ~2E^{ic}(x) E^{kc}(x)-\delta^{ik}[{\vec E}^c(x)\cdot {\vec E}^c(x) -{\vec B}^c(x) \cdot {\vec B}^c(x)]\nonumber \\
&&+\sum_{ q} {\bar \psi}_l(x)[\gamma^k  (i\delta^{lj}{\overrightarrow \partial}^i +gT^c_{lj}A^{ic}(x)) -\gamma^k (\delta^{lj}i{\overleftarrow \partial}^i -gT^c_{lj}A^{ic}(x)) ] \psi_j(x)~]+(antiquarks)
\label{ycp}
\eea
where $A_\nu^c(x)$ is the Yang-Mills potential (color potential) produced by all the quarks plus aniquarks in the system, $(antiquarks)$ means corresponding boundary surface term for antiquarks, ${\bar q}$ means antiquark, $\psi_l(x)$ is the Dirac field of the quark, ${\vec E}^c~({\vec B}^c)$ is the chromo-electric (chromo-magnetic) field and the gauge invariant momentum ${\vec p}_{\rm EM}$ of the Yang-Mills field $A_\nu^c(x)$ and the gauge invariant momentum of the quark are given by
\bea
{\vec p}_{\rm YM}=\int d^3x ~{\vec E}^c(x) \times {\vec B}^c(x),~~~~~~~~~~~~{\vec p}_q = \int d^3x~ \psi^\dagger_l(x) [i\delta^{lj}{\vec \partial} +gT^c_{lj}{\vec A}^c(x)] \psi_j(x).
\label{yeem}
\eea
Note that since the isolated quarks and/or antiquarks are not observed but the quarks and/or antiquarks are confined inside the hadron, whenever we say color field produced by quarks/antiquarks we mean the color field produced by quarks/antquarks inside the hadron. Since we have not observed color outside the hadron there is no color field outside the hadron in our study (see section VI for more details).

Due to confinement the color potential $A_\nu^c(t,r)$ can not decay faster than $r^{-1/2}$ in the Yang-Mills theory or the chromo-electric field ${\vec E}^c(t,r)$ can not decay faster than $r^{-3/2}$ \cite{peym}. Since the chromo-electric field energy density $\frac{{\vec E}^c(t,r) \cdot {\vec E}^c(t,r)}{2}$ can not decay faster than $r^{-3}$ due to confinement the potential energy can not rise slower than ${\rm ln}(r)$. This form of the potential energy is consistent with the fact that the form of the color potential produced by a quark in the classical Yang-Mills theory is not of the Coulomb form $\frac{1}{r}$ \cite{ppn}. This is because if the color potential is of the $\frac{1}{r}$ form then the color charge $q^c$ has to be constant \cite{ppn,pcn} which will predict the $\frac{1}{r}$ form of the
potential energy between quarks and/or antiquarks in a system [similar to the Maxwell theory] which can not explain confinement.

If the potential energy rises linearly (or faster) with distance $r$ then the chromo-electric field ${\vec E}^c(t,r)$ can not decay faster than $r^{-1}$ which means the non-vanishing boundary surface term
\bea
&&\int d^3x ~\partial_k [ ~2E^{ic}(x) E^{kc}(x)-\delta^{ik}[{\vec E}^c(x)\cdot {\vec E}^c(x) -{\vec B}^c(x) \cdot {\vec B}^c(x)]\nonumber \\
&&+ \sum_{ q}{\bar \psi}_l(x)[\gamma^k  (i\delta^{lj}{\overrightarrow \partial}^i +gT^c_{lj}A^{ic}(x)) -\gamma^k (\delta^{lj}i{\overleftarrow \partial}^i -gT^c_{lj}A^{ic}(x)) ] \psi_j(x)~]+(antiquarks)\neq 0\nonumber \\
\label{yvmf}
\eea
at infinite distance. From eqs. (\ref{yvmf}) and (\ref{ycp}) we find
\bea
&&\sum_{ q}\frac{d {\vec p}_q}{dt}+\sum_{\bar q}\frac{d {\vec p}_{\bar q}}{dt}+\frac{d {\vec p}_{YM}}{dt}+\frac{d {\vec p}_{\rm flux}}{dt}=0
\label{ymp}
\eea
where the momentum flux ${\vec p}_{\rm flux}$ is given by
\bea
&&p^i_{\rm flux}=\frac{1}{2}\int d^4x ~\partial_k [ ~2E^{ic}(x) E^{kc}(x)-\delta^{ik}[{\vec E}^c(x)\cdot {\vec E}^c(x) -{\vec B}^c(x) \cdot {\vec B}^c(x)]\nonumber \\
&&+\sum_{ q} {\bar \psi}_l(x)[\gamma^k  (i\delta^{lj}{\overrightarrow \partial}^i +gT^c_{lj}A^{ic}(x)) -\gamma^k (\delta^{lj}i{\overleftarrow \partial}^i -gT^c_{lj}A^{ic}(x)) ] \psi_j(x)~]+(antiquarks)\neq 0.\nonumber \\
\label{ymx}
\eea
In eq. (\ref{ymx}) the time integral $\int dt$ is an indefinite integral but the space integral $\int d^3x$ is definite integral. For the details about the range of the integration $\int d^3x$ in QCD and the size of the hadron see section VI.

Hence extending eq. (\ref{ymp}) from the classical Yang-Mills theory to QCD to study the momentum conservation equation of partons inside the hadron we find
\bea
&&\sum_{ q}<\frac{d {\hat {\vec p}}_q}{dt}>+\sum_{\bar q}<\frac{d {\hat {\vec p}}_{\bar q}}{dt}>+\sum_{ g}<\frac{d {\hat {\vec p}}_g}{dt}>+<\frac{d {\hat {\vec p}}_{\rm flux}}{dt}>=0
\label{qmp}
\eea
where the expectation $<...>$ is defined in eq. (\ref{st}) and the hat means corresponding operator obtained from the classical Yang-Mills theory by replacing $\psi_j,~{\bar \psi}_j,A \rightarrow {\hat \psi}_j,~{\hat {\bar \psi}}_j,~{\hat Q}$ where $Q_\nu^c(x)$ is the (quantum) gluon field. Since we use $A$ for the background field to prove factorization and renormalization in QCD at all orders in coupling constant at high energy colliders \cite{nal}, we have used the notation ${\hat Q}$ for the (quantum) gluon field. Note that the non-perturbative fields ${\hat \psi}_j,~{\hat {\bar \psi}}_j,~{\hat Q}$ in this study are the renormalized fields in the renormalized QCD.

As we saw in section \ref{pbd} since the Yang-Mills theory is obtained by making analogy with the Maxwell theory by extending U(1) gauge group to SU(3) gauge group \cite{pyang,ppn,pcn} we find that as far as the boundary surface at infinite distance is concerned the renormalized non-perturbative QCD at infinite distance predicts the same potential that is predicted by the classical Yang-Mills theory. Hence we find that as far as the boundary surface at infinite distance is concerned a non-vanishing boundary surface term in the classical Yang-Mills theory means a corresponding non-vanishing boundary surface term in the renormalized non-perturbative QCD at infinite distance. The boundary surface term at infinite distance in eq. (\ref{ymx}) in the classical Yang-Mills theory gives non-zero momentum flux ${\vec p}_{\rm flux} \neq 0$. Hence we find that the corresponding boundary surface term at infinite distance in the renormalized non-perturbative QCD gives non-zero momentum flux
\bea
<{\hat {\vec p}}_{\rm flux}> \neq 0.
\label{nz}
\eea
For explicit expression of ${\hat {\vec p}}_{\rm flux}$ in terms of quantum fields ${\hat \psi}_j,~{\hat {\bar \psi}}_j,~{\hat Q}$ in QCD see eq. (\ref{qymx}).

From eq. (\ref{qmp}) and (\ref{nz}) we find
\bea
p_H= \sum_{ q} <{\hat p}_{ q}>+\sum_{\bar q} <{\hat p}_{\bar q}>+\sum_{g} <{\hat p}_{g}>+ <{\hat p}_{\rm flux}>
\label{fnl}
\eea
which reproduces (\ref{fni}).

\section{ Momentum Sum Rule Violation For Parton Distributions of Color Singlet Hadron }

As mentioned in the introduction the hadron state $|H>$ is a color singlet state. Hence one might think that that the confinement cannot violate the momentum sum rule for the parton distributions of color singlet hadrons. This is not true which can be seen as follows.

From eq. (\ref{qmp}) we find that the explicit expressions of $<{\hat {\vec p}}_q>$, $<{\hat {\vec p}}_g>$ and $<{\hat p}^i_{\rm flux}>$ in QCD inside the color singlet hadron $H$ are given by
\bea
&&<{\hat {\vec p}}_g>=<H|{\hat {\vec p}}_g|H>=<H|\int d^3x ~{\hat {\vec E}}^c(x) \times {\hat {\vec B}}^c(x)|H>,\nonumber \\
&&<{\hat {\vec p}}_q>=<H|{\hat {\vec p}}_q|H> =<H| \int d^3x~ {\hat \psi}^\dagger_l(x) [i\delta^{lj}{\vec \partial} +gT^c_{lj}{\hat {\vec Q}}^c(x)] {\hat \psi}_j(x)|H>
\label{qyeem}
\eea
and
\bea
&&<{\hat p}^i_{\rm flux}>=<H|{\hat p}^i_{\rm flux}|H>=<H|\frac{1}{2}\int d^4x ~\partial_k [ ~2{\hat E}^{ic}(x) {\hat E}^{kc}(x)-\delta^{ik}[{\hat {\vec E}}^c(x)\cdot {\hat {\vec E}}^c(x) -{\hat {\vec B}}^c(x) \cdot {\hat {\vec B}}^c(x)]\nonumber \\
&&+\sum_{ q} {\hat {\bar \psi}}_l(x)[\gamma^k  (i\delta^{lj}{\overrightarrow \partial}^i +gT^c_{lj}{\hat Q}^{ic}(x)) -\gamma^k (\delta^{lj}i{\overleftarrow \partial}^i -gT^c_{lj}{\hat Q}^{ic}(x)) ] {\hat \psi}_j(x)~]|H>+(antiquarks)\neq 0\nonumber \\
\label{qymx}
\eea
where $|H>$ is the color singlet (physical) hadron state. Note that in eqs. (\ref{qyeem}) and (\ref{qymx}) the ${\hat {\vec E}}^c(x)$ and ${\hat {\vec B}}^c(x)$ are obtained from
\bea
{\hat F}_{\lambda \mu}^c(x) = \partial_\lambda {\hat Q}_\mu^c(x) - \partial_\mu {\hat Q}_\lambda^c(x) + gf^{cad} {\hat Q}_\lambda^a(x) {\hat Q}_\mu^d(x)
\eea
which contains quantum gluon field ${\hat Q}_\mu^c(x)$ instead of classical Yang-Mills field $A_\mu^c(x)$. Since ${\hat {\vec p}}_g$, ${\hat {\vec p}}_q$ and ${\hat p}^i_{\rm flux}$ are color singlet operators we find that the color singlet hadron state $|H>$ does not make $<H|{\hat {\vec p}}_g|H>$, $<H|{\hat {\vec p}}_q|H>$ and $<H|{\hat p}^i_{\rm flux}|H>$ vanish in eqs. (\ref{qyeem}) and (\ref{qymx}).

Hence we find that that the confinement can violate the momentum sum rule for the parton distributions of color singlet hadrons.

\section{ Finite Size Hadron and Momentum Sum Rule Violation in QCD at High Energy Colliders  }

In our calculation the range of the integration in $\int d^3x$ in eqs. (\ref{qyeem}) and (\ref{qymx}) covers the volume containing the color field. Since we have not observed the color outside the hadron the volume integration in eqs. (\ref{qyeem}) and (\ref{qymx}) in our calculation covers the volume of the hadron.

Hence the volume integration $\int d^3x$ in eqs. (\ref{qyeem}) and (\ref{qymx}) is finite if the size of the hadron is finite. Similarly the volume integration $\int d^3x$ in eqs. (\ref{qyeem}) and (\ref{qymx}) is infinite if one considers the size of the hadron as infinite. If the size of the hadron is finite then the boundary surface term is at finite distance and hence the boundary surface term in eq. (\ref{qymx}) is non-zero. Similarly if one considers the size of the hadron as infinite then the boundary surface term is at infinite distance and hence the boundary surface term in eq. (\ref{qymx}) is non-zero which we have shown in eq. (\ref{nz}). Note that as shown above the boundary surface term at infinite distance in eq. (\ref{nz}) is non-zero if the confinement potential energy at large distance $r$ can not rise slower than ${\rm ln}(r)$ whereas boundary surface term at finite distance in eq. (\ref{qymx}) is non-zero for any form of the confinement potential energy. Hence we find from eqs. (\ref{qymx}) and (\ref{nz}) that the boundary surface term $<{\hat p}^i_{\rm flux}>$ in QCD is non-zero irrespective of whether the size of the hadron is finite or infinite.

The non-zero value of $<{\hat p}^i_{\rm flux}>$ in eq. (\ref{qymx}) has to be calculated by using non-perturbative QCD similar to the non-zero values of $<{\hat {\vec p}}_g>$ and $<{\hat {\vec p}}_q>$ in eq. (\ref{qyeem}) which have to be calculated by using non-perturbative QCD.

Although we have considered the infinite size of the hadron in eq. (\ref{nz}) to demonstrate that, unlike QED where the boundary surface term vanishes at infinite distance, the boundary surface term in QCD does not vanish at infinite distance. But in reality the size of the hadron is finite which means, as discussed above, the boundary surface is at finite distance and hence the boundary surface term at finite distance is non-zero for any form of the confinement potential energy. Hence we find that the momentum sum rule in QCD at high energy colliders obtained from the first principle as given by eq. (\ref{fnl}) must include the non-zero boundary surface term $<{\hat p}_{\rm flux}>$ if one wants to describe the hadron from quarks and gluons and if one uses the QCD as the fundamental theory of the nature describing the interaction between quarks and gluons.

\section{Conclusions}
Momentum sum rule in QCD is widely used at high energy colliders. Although the exact form of the confinement potential energy is not known but the confinement potential energy at large distance $r$ can not rise slower than ${\rm ln}(r)$. In this paper we have found that if the confinement potential energy at large distance $r$ rises linearly with $r$ (or faster) then the momentum sum rule in QCD is violated at the high energy colliders if we consider the hadron size as infinite. For finite size hadron we have found that the momentum sum rule in QCD is violated at the high energy colliders for any form of the confinement potential energy.

\end{document}